# Data Migration among Different Clouds

Ismail Hababeh

Ismail.Hababeh@gju.edu.jo
Faculty of Computer Engineering and Information Technology, German-Jordanian University, Jordan

*Abstract*—Cloud computing services are becoming more and more popular. However, the high concentration of data and services on the clouds make them attractive targets for various security attacks, including DoS, data theft, and privacy attacks. Additionally, cloud providers may fail to comply with service level agreement in terms of performance, availability, and security guarantees. Moreover, users may choose to utilize public cloud services from multiple vendors for various reasons including fault tolerance and availability. Therefore, it is of paramount importance to have secure and efficient mechanisms that enable users to transparently copy and move their data from one provider to another. In this paper, we explore the state-of-the-art inter-cloud migration techniques and identify the potential security threats in the scope of Hadoop Distributed File System HDFS. We propose an inter-cloud data migration mechanism that offers better security guarantees and faster response time for migrating large scale data files in cloud database management systems. The proposed approach enhances the data security processes used to achieve secure data migration between cloud nodes thus improves applications' response time and throughput. The performance of the proposed approach is validated by measuring its impact on response time and throughput, and comparing the performance to that of other techniques in the literature. The results show that our approach significantly improves the performance of HDFS and outperforms its counterparts.

## I. Introduction

Public cloud computing paradigm is continuously gaining momentum as a reliable and cost effective solution to public computing needs. It delivers computing resources over the Internet and stores users' data on the cloud. Customers enjoy unlimited data storage and computation power as per pay bases while the underlying hardware/software resources are maintained and managed by the Cloud Service Provider (CSP). The cloud storage model can be classified into private and public clouds. Private clouds provide services and utilities to a small group of enterprises. The data and service management of the cloud is performed locally by the enterprise. In public clouds, organizations and individuals have cloud services from a third party who manages the cloud [4]-[7]. Most of enterprises prefer to deploy their own private clouds for data storage driven by the privacy and security concerns [30]. However, the world is producing very vast amount of data (big data) in every day routine. The traditional data processing and storage systems cannot handle the characteristics and the requirements of big data and no longer suitable to meet the new big data challenges [16][17]. Therefore, outsourcing data and computation to public clouds becomes a more practical solution. The Apache Hadoop [9] is an open source cloud database solution developed to address the unique challenges of big data. Hadoop has its own file system called Hadoop Distributed File System (HDFS). HDFS architecture was initially designed to store and process data assuming HDFS is operating in a secure world. However, HDFS architecture has very critical problems which lead to loss of sensitive information. A fix for HDFS has been proposed in [2] to provide secure management of data in HDFS.

As the demand for cloud services increases, the competition between cloud service providers also increases. For cloud clients to benefit from such competition, they should be able to freely and easily migrate their data from one cloud to another. This migration could be triggered, for example, by cost, performance, bandwidth, reliability, security, or legal issues. However, this migration may incur potential threats to the privacy of the clients and the security of their data [32]. In this work, we propose a secure inter-cloud data migration architecture that takes into consideration the efficiency of the migration process, the privacy of the clients, and the confidentiality of the data. Our proposed protocol provides secure and efficient inter-cloud data migration on Hadoop Distributed File System (HDFS) based clouds. The internal data security is ensured by the HDFS security layer [2]. Our protocol requires that sensitive data are always stored in encrypted format with a key known only to the data owner. This requirement greatly enhances the data migration efficiency as the migration engine would not be responsible for encrypting and decrypting large chunks of data while being migrated [20][31]. More importantly, this helps to boost the security of the migrated data against both insiders and outsiders. Data owner encryption ensures data confidentiality while our data migration protocol ensures data integrity [29]. Moreover, our protocol protects against attacks that try to utilize the migration process to cause data loss. The acknowledgment mechanism that we introduce ensures that data at the source is only removed after the destination successfully received it. The security and performance evaluation of our protocol shows its superiority over the state-of-the-art inter-cloud data migration protocol [3].

Inter-cloud data migration is a relatively new problem and its security guarantees has not received the due attention. The inter-cloud secure data migration has also been discussed in [3]. However, the work in [3] has some very serious security and performance issues that we will thoroughly discuss in Section II.

The rest of the paper is organized as follows. In Section II, we go over the background material and the related work. In Section III, we present our secure inter-cloud data migration protocol. In Section IV, we present the experimental results. In Section V, we discuss the performance evaluation. Finally, in Section VI, we draw the conclusion and future work.

## II. RELATED WORK

The Apache Hadoop is proposed and implemented to deal with big data requirements. The Hadoop storage file system is Hadoop Distributed File System (HDFS) which is distributed, scalable and portable file system. The majority of cloud service providers use HDFS for cloud storage to utilize the scalable and reliable distributed storage features of Hadoop [9]. HDFS is comprised of Hadoop clusters. Each cluster contains one Name Node and multiple Data Nodes [1]. The Name Node stores metadata related to files and directories such as size, permissions, type, modifications, etc. In addition, it stores information related to clients' data on Data Nodes. The actual data is stored on Data Nodes. The client requests are processed using Map-Reduce parallel programming model [11]. The client submits read or write requests to Map-Reduce node called Job Tracker. The Job Tracker performs HDFS operations on client's behalf. The Job Tracker divides client's request into different tasks to release burden at Job Tracker. The Job Tracker launches tasks which run on nodes called Task Trackers. Each Task Tracker then communicates with HDFS Name Node and Data Nodes on behalf of the user [2][14]. Like other file systems, we can perform different I/O operations on HDFS for read, write and update. In order to read or write to HDFS, client first contacts the Name Node to get the block IDs for the Data Nodes and then submits his requested job to the Job Tracker along with the block IDs. The Job tracker launches one or more Task Trackers which perform the requested operations on data nodes and returns the results to the client.

The original implementation of HDFS is not designed with security in mind and has many security issues, including: the lack of access control mechanisms, especially at the level of Name Node and Data Nodes and the lack of data confidentiality guarantees as data is stored in plain-text format. The authors in [2] propose a Kerberos-based [11] secure authentication model for HDFS. It allows only authorized users to access HDFS, a user can only access his own files, and a user can only modify his own Map-Reduce jobs. Kerberos is a trusted third-party protocol that is used to securely prove identity of nodes to each other over insecure networks. Further communications among the different modules of HDFS are totally based on delegation tokens generated by the Name Node after correctly authenticating the user using Kerberos. In [19], the authors present a data encryption mechanism for data nodes. The proposed mechanism guards the encryption key from insiders by separating the public and private keys on the client and the data node. These mechanisms greatly enhance the intra-HDFS security; however, they do not address the inter-HDFS security concerns.

The first attempt to address the inter–HDFS security concerns was by O'Malley et al. [3]. The authors discuss the secure data migration between cloud storages that are based on Hadoop distributed file system. The source cloud initiates the data migration based on users' request to a target cloud. The data migration process is triggered by the user through a request to the source cloud. The source cloud authenticates the user and validates his authorization for the requested data migration. After successful authorization of the user, the source cloud initiates a SSL connection with the target cloud, after which, the secure inter-cloud data migration starts. We perform thorough analysis of this protocol and found that it contains noticeable security concerns and performance issues. A recent related study [18] also discusses the security of storage data migration between different clouds. Both [3] and [18] address propose secure migration mechanisms that depend on secure socket layer (SSL) connection between the name nodes of both source and target cloud systems. Target name node generates a temporary session key (T.Kdn), a random number (R.hash) and series of tickets encrypted by T.Kdst to communication with the source nodes, compute the double hash value and return the encrypted tickets to the source name node respectively. However, they didn't specify how the user is authenticated at both source and destination clouds. The target cloud should verify whether the user has truly requested the source cloud to initiate data migration on his behalf rather than a malicious away to illegally exfiltrate data. Moreover, the acknowledgment model has not been correctly defined. This may present serious security issues. On one hand, if no acknowledgment is assumed after sending the data from the source to the target cloud, the source Data Node deletes the data after being transferred without any knowledge whether it has been correctly received. This enables a man-in-the-middle attacker to create considerable data lost. The attacker drops the packets, the source deletes them, and the receiver did not get any data. On the other hand, if the target acknowledges the successful reception of the packets, the administrator would be vulnerable to flood attacks over duplicate reception of packets carrying the same ticket. Finally, these mechanisms cause unnecessary network bandwidth and processing overhead due to the extra transmission of tickets and extra encryptions as we show in Section IV. For example, it is not clear why double hash is being used.

In [21], the authors propose a simple approach to encode and decode actual data during migration of large databases. However, synchronization has to be performed as multiple users can access the server simultaneously. In case of server failure, the receiver will not be able to communicate with the sender; a recovery mechanism is required instead. Authors of [22] propose a privacy-preserving architecture for inter-Cloud data sharing. This architecture supports a proactive mechanism which relies on network services, like e-mail, to send asynchronous notification messages to all clients of the data share group, with the link to that information into the STaaS provider infrastructure. However, it is vulnerable to identity masquerading, through for example, compromise of the client's email. Authors of [23] propose an extension to the ISO 27001:2005 standard and the evaluation of ISO 27001:2005 completeness towards cloud security. They include a control objective for virtualization management, with two controls: virtualization and virtual machines control. They propose that the information involved in virtual machines is protected from internal and external threats, and secured in transit. However, neither the information nor management and controlled involved in virtual machines are well defined. The mechanism of securing inter-cloud data is not provided as well.

In [24], the authors propose an Inter-cloud Resource Integration System (Iris) using nested virtualization and OpenFlow technologies. Iris dynamically configures and provides a virtual infrastructure over inter-cloud resources, on which an IaaS cloud can run. The inter-cloud federation system establishes secure isolation between HaaS and IaaS to avoid security incidents in a HaaS data center. HaaS system enables migrant VMs between data centers using the same IaaS API. However, during VM migration, the memory pages are copied over HaaS data centers which introduce heavy inter-cloud communication traffic after migration. Therefore, the migration time grows as the network latency increases. Moreover, Iris does not support migration of VMs with different security levels over the hybrid clouds which allow for high risk attacks and large number of threats. Authors of [25] present a privacy-aware VM migration framework, which causes minimal service level agreement (SLA) disruption. All the VMs are divided into two categories according to the data sensitivity: sensitive VMs and non-sensitive VMs. For consideration of data privacy, the sensitive VMs should run in private cloud. However, VM migration should only be allowed if both source and destination servers are trusted which is not the case between private cloud and public cloud. Moreover, threats can take place through migrating sensitive data from private cloud to public cloud as no authorized data migration mechanism is applied between source and destination clouds. The problem of secure data migration among clouds can be mapped and benefit from problems in other domains such as sensor network security [33]-[43] and network coding [44][45].

### III. SECURE INTER-CLOUD DATA MIGRATION

We present here a complete inter-cloud secure data migration protocol that ensures the integrity and the confidentiality of user's data while preserving user's privacy. Our protocol builds on the intra-Hadoop security protocol [2] and addresses all the security concerns of the current state-of-the-art inter-cloud data migration protocol [3]. Also, our protocol addresses the performance concerns of the current inter-cloud data migration protocol by reducing its networking and processing overhead. In the following, we present the assumptions, the initial setup and the detailed protocol steps.

*A. Assumption*

We assume that a user (U) who plans to move his data from a source cloud (SC) to a target cloud (TC) has already established user accounts with both the source cloud and the destination cloud. We assume that SC and TC are trusted by the user, but TC and SC may not trust each other.

*B. Initial setup*

The user initiates the data migration process by generating a symmetric key $K_t$. The user then uses his secure communication channels with both the source cloud and the target cloud (ID/Password pairs) to deliver the key to both the source cloud and the target cloud. This step ensures that no one but the legitimate owner of the data can initiate the migration. Figure 1 presents the user authentication steps at both source and target clouds which are described as follows:

1. The user login to the source and the target clouds independently using his login credentials.
2. The user generates a random key $K_t$
3. The key $K_t$ is encrypted by user's account password at source and is sent to source.
4. The key $K_t$ is encrypted by user's account password at target and is sent to the target.

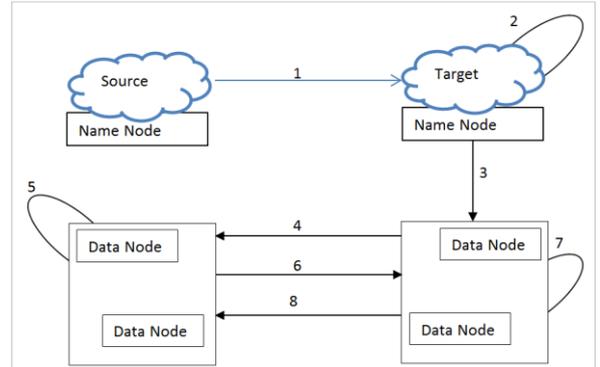

Fig. 1. User authentication at the source and the destination clouds

The user then executes the following steps that are illustrated in figure 2 to finalize the data migration from the source to the target cloud:

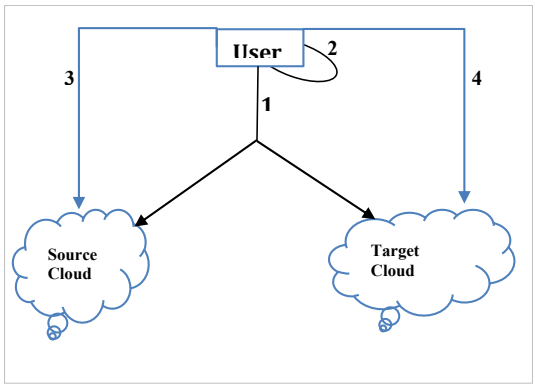

Figure 1: Inter-Cloud data migration protocol

1. The source cloud sends to the target cloud the necessary metadata of the user, such as data block IDs, Data Node addresses and any other related information that locates user's data on the Data Nodes of the source cloud. This metadata is encrypted using $K_t$.
2. The target generates block access tokens and encrypts them using $K_t$.
3. The target shares these block access tokens with its data nodes.
4. The target Data Node requests for reading data from the source Data Node and sends them the respective token.
5. The source Data Node receives the request and decrypts the token to verify authenticity of the request.
6. The source Data Node sends the data to the target Data Node and also sends the computed hash value of data encrypted by $K_t$. The source Data Node starts a timer and waits for acknowledgment. If acknowledgment is not received in time due to network problems or any other issues, the packet is retransmitted. The source Data Node keeps retransmitting until either a successful acknowledgment is received or a predefined maximum number of retransmissions (MaxRet) is reached. In the latter case, the administrator in the source is prompted.
7. The target Data Node receives the data and verifies its hash value.
8. If correctly verified, the target Data Node sends acknowledgment back to the source Data Node encrypted by $K_t$. If the acknowledgment is lost due to network problems or some other issue, the target Data Node may receive more than one copy of the same packet due to retransmissions. In this case all the duplicate copies are dropped. However, if the number of duplicate copies exceeds MaxRet, the administrator in the target is prompted.
9. The source data node receives acknowledgement and deletes the successfully delivered data.

### IV. EXPERIMENTAL RESULTS

Considerable attentions have been made for securing inter-cloud data migration to improve HDFS performance. In this context, several approaches make modifications in how data is encrypted, and transferred between source and target cloud nodes.

Indeed, most studies in the literature present extra security overhead [2][3] which influence the system performance especially when big data files are distributed over large number of cloud data nodes.

The demand for efficient approach to such data migration security facilities has been addressed in our approach, concerning data security, privacy, and integrity within the framework. Accordingly, a secure big data migration system over heterogeneous inter-cloud environment has been developed. To demonstrate the feasibility of the proposed system, extensive experiments are conducted to evaluate the performance of our approach for migrating big data files. The cloud system performance is tested on big data file sizes (100 MB, 1GB, 2GB, 4GB, 8GB, and 16GB) and measure the migration time overhead.

To setup the cloud system of such experiments, Apache Hadoop 2.6.0 [26] is installed on PowerEdge R720 servers that consists of 2 x (6 Core) Intel Xeon E5-2630 v2 @ 2.60GHz, 64.0 GB RAM, runs Linux Ubuntu 10.04, java 1.0.7- openjdk and connected with Gigabit Ethernet NICs.

PuTTY 0.64 [27] is installed to configure the experiments and compare the performance of our protocol against the baseline Hadoop, the secure intra-cloud Hadoop [2], and the secure intra-cloud Hadoop [3]. User authentication in both source and target clouds are done using a private key through Putty SSH authentication system. To start data migration process, the user private key is integrated with a public key. After passing the authentication processes, the user is connected to the target cloud and startup the data migration process. Figure 4 depicts the time (sec.) required to migrate 100MB data file by using HDFS baseline.

Table I presents the data migration time (sec.) required to authenticate, extract and load 6 different big data file sizes using HDFS Baseline, secured intra, inter and proposed methods at migration rate of 64Mb/s.

TABLE I. DATA MIGRATION TIME (SEC.) FOR BASELINE AND SECURED HDFS METHODS

| Method | 100MB | 1GB | 2GB | 4GB | 8GB | 16GB |
|---|---|---|---|---|---|---|
| Baseline | 12.5 | 128.3 | 262.5 | 550.2 | 1156.1 | 2392.9 |
| Secured HDFS[2] | 14.1 | 170.4 | 381.9 | 814.8 | 1761.7 | 3709.6 |
| Secured HDFS[3] | 13.9 | 161.8 | 348.4 | 741.9 | 1601.4 | 3363.1 |
| Proposed secured HDFS | 12.7 | 145.8 | 302.1 | 636.9 | 1356.7 | 2843.2 |

This table shows that the data migration results in our method outperform other methods in comparison and much closer to the baseline case. If we assume that the extract and load times are same in all methods, then the securing processes in our method consumes less time and hence improves the cloud system performance.

The delay time caused by secured data migration processes is defined as the difference between the data migration time in secured HDFS (DMTS) and the data migration time in the HDFS baseline (DMTB). Equation 1 illustrates the calculated formula of the delay time.

$$\text{Delay Time} = \text{DMTS} - \text{DMTB} \quad (1)$$

Figure 5 shows the delay time (sec.) caused by secured HDFS methods.

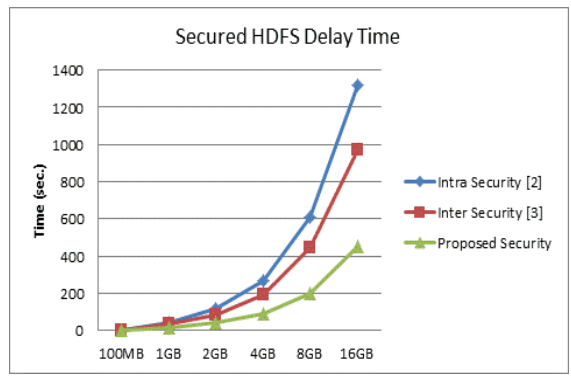

Fig. 2. Data migration delay time in secured HDFS

Based on our data migration experiments, it can be inferred that the delay time caused by our approach is much less than other methods which leads to achieve high performance. However, applying the secured data migration processes will generally decrease HDFS performance as it cause delay time.

The HDFS performance degradation is defined as the delay time caused by secured HDFS method divided by the data migration time in HDFS baseline.

Table II shows the performance degradation caused by secured HDFS approaches.

TABLE II. PERFORMANCE DEGRADATION CAUSED BY SECURED HDFS METHODS

| Method | 100MB | 1GB | 2GB | 4GB | 8GB | 16GB |
|---|---|---|---|---|---|---|
| Secured HDFS [2] | 12.8% | 32.8% | 45.5% | 48.1% | 52.4% | 55.0% |
| Secured HDFS [3] | 9.9% | 26.1% | 32.7% | 34.8% | 38.5% | 40.5% |
| Our secured HDFS | 1.4% | 13.6% | 15.1% | 15.8% | 17.4% | 18.8% |

During inter-cloud data migration, the secured HDFS memory pages are copied over cloud data nodes. Therefore, the secured data migration time increases as the network latency increases. Despite the fact that the secured HDFS methods are bit slower than general HDFS, the experiments results show that inter-cloud data migration time are nearly close regardless the file size. However, the results emphasize that our approach has marginally less overhead compared to other approaches [2][3].

V. SECURITY ANALYSIS

The migration process mainly consists of three communication phases:
(i) The setup communication among the user, the source and the target clouds
(ii) The communication for address sharing between the source and the target cloud
(iii) The communication between the source data nodes and

the target data nodes for actual data transfer.

The communication between the user and the source/target cloud is protected by the shared credentials between the user and the source/target cloud. The communication between the source and the target name nodes for metadata sharing is protected by Kt. The attacker might capture the metadata transmitted from the source to the target, however, the contents cannot be read, which makes our solution secure from passive attackers. The encryption of the metadata also helps to guard against some active attackers. Attacker may present himself to the source as the target or vice versa, but he cannot encrypt/decrypt the necessary metadata and thus the attack fails. In the phase of communication between the source data node and the target data node, the target data node initially sends the encrypted block access tokens to the source data node where it can be captured by attackers but cannot be decrypted. The source data node decrypts the block access tokens and sends the user's data along with hash value of the data concatenated with a random nonce to prevent replay attacks. The source only encrypts the hash value because user's data is already stored in encrypted form. One possible attack here is to change the data packets during transmission or block them in the middle. The integrity of the data is verified through the associated hash value and the successful reception of the data is verified by the returned acknowledgment. The integrity of the acknowledgment is also guaranteed.

On the other hand, as we mentioned earlier, the inter-cloud data migration protocol presented in [3] lacks proper authentication of the user with both the source and the target clouds. Also, the protocol wrongly assumes full trust between the source and the target clouds. The source and target may not fully trust each other. Also, a corrupted insider at the source may start migration of data without the user's consent or request. The target cloud does not verify the proper initiation of the migration by the right user. On the other hand, our proposed protocol ensures that the migration is only triggered by the user and provides both the target cloud and the source cloud the means to correctly verify the process.

Next, we discuss the behavior of our protocol in the face of the man-in-the-middle attack. Assume that a man-in-the-middle attacker controls the communication in step 4. Also assume that the goal of the attacker is to divert the data being migrated from the legitimate target to his own storage in a bid to steal the data. The attacker stops the message from the target data node and presents himself as the target data node. The source data node correctly verifies the token and sends the data along with the hash value back to attacker. This attack will fail because the hash value is encrypted by Kt which is only known to the legitimate target Data node. Also the data itself is encrypted by a key which is only known to the legitimate user. Therefore, even though the attacker may get the data he cannot benefit from it. If the goal of the attacker is to cause data loss by making the source to delete the data without being successfully received and stored in the target. As we mentioned earlier, the source cloud only deletes the data when it receives the acknowledgment from the related target data node. The acknowledgment message is computed using the hash value sent by the source and Kt. The attacker cannot forge a correct acknowledgment as he does not possess Kt and, therefore, would not be able to convince the source that the data has been successfully received unless it is truly received by the legitimate target Data node. Table III summarizes the security status of each step of our protocol.

TABLE III. SECURITY STATUS IN OUR PROTOCOL STEPS

| Step | Evaluation |
|---|---|
| 1 | The communication is encrypted by Kt. |
| 3 | The tokens are shared using internal security of HDFS |
| 4 | The tokens are encrypted using $K_t$. |
| 6 | Data is encrypted by a key which is only known to the user and integrity is guarded by a hash encrypted by $K_t$ |
| 8 | The acknowledgment is encrypted by Kt. |

Our protocol addresses all the security issues of [3]. The migration process requires that the target cloud should know the necessary metadata at the source name node. Our protocol achieves this goal by using the common key Kt (established by the user between the source and the target clouds). Moreover, this key serves as an authentication to both the source and the target clouds that the migration has been triggered by the owner of the data. Finally, having this common key enables both the target and the source clouds to generate and verify the necessary tokens without having them to be unnecessarily transferred from one cloud to another.

Our inter-cloud data migration does not only provide better security guarantees but also it slightly decreases the network and processing overhead compared to [3]. Our protocol optimizes the communications and the encryption and decryption operations between the source and the target clouds. An important factor that we have introduced is that the data does not need to be encrypted or decrypted during the migration at the source and the target cloud. If the data is sensitive, then it would have been stored in encrypted format. Otherwise integrity check would be sufficient during the migration when we deal with public clouds.

## VI. CONCLUSION AND FUTURE WORK

In this work, we propose and evaluate a new inter-cloud secure data migration protocol. The protocol addresses the security concerns of the current state-of-the-art secure inter-cloud data migration protocols. The new protocol ensures data integrity and confidentiality during the migration process. The protocol uses secure mutual authentication between the source and the target clouds to countermeasures potential sabotage attacks that may use the migration process to destroy the data being migrated. The protocol also ensures that the migration process is initiated by the legitimate owner of the data and not by a malicious perpetrator trying to ex-filtrate the data. The contribution of our protocol lies in the extra security guarantees provided with even (marginal) less performance overhead compared the state-of-the-art inter-cloud data migration protocols.

We perform both external and internal evaluation of our approach. In the internal evaluation, we measure the impact of using our technique on inter-cloud services performance like elapsed time caused by securing HDFS and throughput. In the external evaluation, we compare the performance of our approach to that of other techniques in the literature.

The results show that our approach significantly improves services requirement satisfaction in cloud systems. This conclusion requires more investigation and experiments. Therefore, as future work we plan to investigate our approach on larger data files distributed over large number of cloud data nodes. In addition, we will consider applying search based technique to perform more intelligent data reallocation. Finally, we intend to filter the data into sensitive data where security concerns have high priority to be addressed and non-sensitive data where security concerns have less priority.

REFERENCES

[1] Konstantin Shvachko, Hairong Kuang, Sanjay Radia, Robert Chansler. The Hadoop Distributed File System. In Proceedings of the 26$^{th}$ IEEE Symposium on Mass Storage Systems and Technologies, 3-7 May 2010, pp. 1-10.

[2] Owen O'Malley, Kan Zhang, Sanjay Radia, Ram Marti, and Christopher Harrell. Hadoop Security Design. Technical Report, 10/2009.

[3] Qingni Shen; Lizhe Zhang, Xin Yang, Yahui Yang, Zhonghai Wu, Ying Zhang, "SecDM: Securing Data Migration between Cloud Storage Systems," Dependable, Autonomic and Secure Computing (DASC), 2011 IEEE Ninth International Conference on , vol., no., pp.636,641, 12-14 Dec. 2011

[4] Michael Armbrust, Armando Fox, Rean Griffith, Anthony D. Joseph, Randy Katz, Andy Konwinski, Gunho Lee, David Patterson, Ariel Rabkin, Ion Stoica, and Matei Zaharia, "A view of cloud computing," in 2010 Communications of the ACM, April 2010, vol-53, pp-50-58

[5] S Carlin, K Curran, "Cloud Computing Security," in 2009 International Journal of Ambient Computing and Intelligence.

[6] F. Ahmed, S. Bouktif, A. Serhani and I. Khalil, "Integrating function point project information for improving the accuracy of effort estimation" Proceedings of the Second International Conference on Advanced Engineering Computing and Applications in Sciences (ADVCOMP'08), Valencia, Spain, September 29 – October 4 2008, pp. 193-198.

[7] S. Bouktif, F. Ahmed, I. Khalil and G. Antoniol, "A novel composite model approach to improve software quality prediction," Information and Software Technology (an Elsevier journal), Volume 52, Issue 12, pp. 1298-1311, December 2010.

[8] http://hortonworks.com/wp-content/uploads/2012/06/Apache-Hadoop-Big-Data-Refinery-WP.pdf. Last accessed on April 8$^{th}$, 2015.

[9] http://hadoop.apache.org/ Last accessed on April 8$^{th}$, 2015.

[10] http://www-01.ibm.com/software/data/bigdata/. Last accessed on April 8$^{th}$, 2015.

[11] Jennifer G. Steiner, Clifford Neuman, Jeffrey I. Schiller, "Kerberos: An Authentication Service for Open Network Systems".

[12] J. Dean, S. Ghemawat, "MapReduce: Simplified Data Processing on Large Cluster," in 2004 Proc. Of the 6th Symposium on Operating Systems Design & Implementation, San Francisco-CA.

[13] Cong Wang; Qian Wang; Kui Ren; Wenjing Lou, "Privacy-Preserving Public Auditing for Data Storage Security in Cloud Computing," in 2010 Proceedings IEEE INFOCOM , vol., no., pp.1,9, 14-19 March 2010.

[14] Dhruba Borthakur , "The Hadoop Distributed File System: Architecture and Design".

[15] Ali Khajeh-Hosseini, David Greenwood, Ian Sommerville,"Cloud Migration: A Case Study of Migrating an Enterprise IT System to IaaS" in 2010 IEEE 3$^{rd}$ International Conference on Cloud Computing CLOUD, pp. 450-457

[16] Goss, Raymond Gardiner; Veeramuthu, Kousikan, "Heading towards big data building a better data warehouse for more data, more speed, and more users," in 2013 24th Annual SEMI , Advanced Semiconductor Manufacturing Conference (ASMC) , vol., no., pp.220,225, 14-16 May 2013

[17] Steve LaValle, Eric Lesser, Rebecca Shockley, Michael S. Hopkins , Nina Kruschwitz, "Big Data, Analytics and the Path From Insights to Value," in MIT Solan Management Review, Vol. 52 No.2.

[18] Rakesh Sachdeva, Prabhpreet Kaur "SSM: Secure Storage Migration among Cloud Providers" in The International Journal Of Science & Technoledge. Vol 2, issue 3, March 2014, pp. 120-126.

[19] Zhonghan Cheng, Diming Zhang, Hao Huang, Zhenjiang Qian, "Design and Implementation of Data Encryptionin Cloud based on HDFS". International Workshop on Cloud Computing and Information Security (CCIS 2013). PP 274-277.

[20] I. Khalil and S. Bagchi, "SECOS: Key Management for Scalable and Energy Efficient Crypto On Sensors," Proceedings of IEEE Dependable Systems & Networks (DSN'06), June, 2006.

[21] R.Vinodha, R.Suresh, "Secure Migration of Various Database over A Cross Platform Environment". International Journal Of Engineering And Computer Science, Vol. 2 - Issue 4, April, 2013. pp. 1072 -1076.

[22] Antonio Famulari, Francesco Longo, Giuseppe Campobello, Thomas Bonald, Marco Scarpa, "A Simple Architecture for Secure and Private Data Sharing Solutions". 2014 IEEE Symposium on Computers and Communication (ISCC), 23-26 June 2014. PP 1-6.

[23] Sasko Ristov, Marjan Gusev, Magdalena Kostoska, "A New Methodology for Security Evaluation in

[24] Cloud Computing", MIPRO, 2012 Proceedings of the 35th International Convention, IEEE, 21-25 May, 2012. PP 1484-1489.

[25] Ryousei Takano, Atsuko Takefusa, Hidemoto Nakada, Seiya Yanagita, Tomohiro Kudoh, "Iris: An Inter-cloud Resource Integration System for Elastic Cloud Data Centers". CLOSER 2014 - 4th International Conference on Cloud Computing and Services Science. pp 103-111.

[26] Hongli Zhang, Panpan Li, Zhigang Zhou, Junchao Wu, Xiangzhan Yu, "A Privacy-aware Virtual Machine Migration Framework on Hybrid Clouds". Journal of Networks, Vol. 9, no. 5, MAY 2014. PP. 1086-1096.

[27] https://hadoop.apache.org/ Last accessed on September 8$^{th}$, 2015.

[28] http://www.chiark.greenend.org.uk/~sgtatham/putty/. Last accessed on April 8th, 2015.
[29] I Khalil, S Bagchi, N Shroff "Analysis and evaluation of SECOS, a protocol for energy efficient and secure communication in sensor networks," Ad Hoc Networks 5 (3), 2007. pp. 360-391
[30] IM Khalil, A Khreishah, S Bouktif, A Ahmad, "Security concerns in cloud computing," Tenth International Conference on Information Technology: New Generations (ITNG), 2013, pp. 411-416
[31] I Khalil, A Khreishah, M Azeem, "Consolidated Identity Management System for secure mobile cloud computing," Computer Networks, Vol. 65, 2014, pp. 99-110
[32] IM Khalil, A Khreishah, M Azeem, "Cloud computing security: A survey," Computers 3 (1), 2014, pp. 1-35.
[33] R. K. Panta, S. Bagchi and I. Khalil, "Efficient wireless reprogramming through reduced bandwidth usage and opportunistic sleeping," Ad Hoc Networks, Volume 7, Issue 1, January 2009, pp. 42-62.
[34] I. Khalil, "ELMO: Energy aware local monitoring in sensor networks," IEEE Transactions on Dependable and Secure Computing, Volume 8, Issue 4, pp. 523-536, 2011.
[35] M. Hayajneh, I. Khalil and Y. Gadallah, "An OFDMA-based MAC protocol for under water acoustic wireless sensor network," Proceedings of the 2009 ACM International Conference on Wireless Communications and Mobile Computing (IWCMC'09), Leipzig, Germany, June 21 – 24 2009, pp. 810-814.
[36] I. Khalil, "MCC: Mitigating colluding collision attacks in wireless sensor networks," Proceedings of the IEEE Global Communications Conference (IEEE GLOBECOM'10), December 6 – 10, 2010, Miami, Florida, USA, pp. 1-5.
[37] I. Khalil, "MIMI: Mitigating packet misrouting in locally-monitored multi-hop wireless ad hoc networks," Proceeding of the IEEE Global Communications Conference (IEEE GLOBECOM'08), New Orleans, USA, November 30 – December 4 2008, pp. 1-5.
[38] I. Khalil, "MPC: Mitigating stealthy power control attacks in wireless ad hoc networks" Proceedings of the IEEE Global Communications Conference (IEEE GLOBECOM'09), Nov. 30 – Dec. 4, 2009, Honolulu, Hawaii, USA, pp. 1-6.
[39] . Khalil, M. Hayajneh and M. Awad "SVNM: Secure verification of neighborhood membership in static multi-hop wireless networks," Proceedings of the IEEE Symposium on Computers and Communications (ISCC'09), Sousse, Tunisia, July 5-8 2009, pp. 368-373.
[40] I. Khalil, M. Awad, S. Bouktif, and F. Awwad, "MSN: Mutual secure neighbor verification in multi-hop wireless networks" Security and Communication Networks, Volume 5, Issue 2, pp. 186-196, February 2012.
[41] I. Khalil, M. Awad and A. Khreishah, "CTAC: Control traffic tunneling attacks' countermeasures in mobile wireless networks," Computer Networks. Volume 56, Issue 14, pp. 3300–3317, September 2012.
[42] I. Khalil, A. Khreishah, F. Ahmed, and K. Shuaib, "Dependable Wireless Sensor Networks for Reliable and Secure Humanitarian Relief Applications," Ad Hoc Networks, Volume 13, pp. 94-106, 2014.
[43] I. Hababeh, I. Khalil, A. Khreishah and S. Bataineh, "Performance Evaluation of Wormhole Security Approaches for Ad-Hoc Networks," Journal of Computer Science, Volume 9, Issue 12, pp. 1626-1637, November 2013.
[44] A. Khreishah, I. Khalil, and J. Wu, "Universal Opportunistic Routing Scheme using Network Coding," Proceedings of the 9th Annual IEEE Communications Society Conference on Sensor, Mesh and Ad Hoc Communications and Networks (SECON), June 18-21, 2012, Seoul, Korea, pp. 353-361.
[45] A. Gharaibeh, A. Khreishah, I. Khalil, and J. Wu, "Asymptotically-Optimal Incentive-Based En-Route Caching Scheme," Proceedings of the 11th IEEE International Conference on Mobile Ad hoc and Sensor Systems (IEEE MASS 2014), October 27-30, 2014, Philadelphia, Pennsylvania, USA.